\documentclass[aps,prc,twocolumn,nofootinbib]{revtex4-2}
\usepackage[T1]{fontenc}
\usepackage{times}
\usepackage{amsmath,amssymb}
\usepackage{hyperref}  
\usepackage{graphicx}
\usepackage{subcaption}
\usepackage[dvipsnames]{xcolor}
\usepackage{caption}  
\renewcommand{\d}{\mbox{d}}

\begin{document}

\title{Demystifying the fusion mechanism in heavy-ion collisions: \\
  A six-dimensional Langevin dissipative dynamics approach}

\author{Y. Jaganathen$^{1}$}
\author{M. Kowal$^{1}$}
\email{michal.kowal@ncbj.gov.pl}  
\author{K. Pomorski$^{1}$}

\affiliation{$^{1}$National Centre for Nuclear Research, Pasteura 7, 02-093 Warsaw, Poland}

\date{\today}  

\begin{abstract}
We present an in-depth investigation of heavy-ion fusion dynamics using a six-dimensional Langevin framework that enables unrestricted motion of the asymmetry parameter. The stochastic formalism naturally incorporates friction effects and energy fluctuations, providing a detailed understanding of the fusion process. The dynamics transition into the overdamped regime, facilitating rapid neck stabilization while effectively capturing the interplay between shape and rotational degrees of freedom. This approach achieves excellent agreement with experimental spin distributions and fusion cross-sections, establishing a robust foundation for forthcoming studies on the synthesis of superheavy elements and the exploration of the enigmatic fusion hindrance mechanism.

\vspace{3pt}

\textbf{Note:} This is the authors' manuscript published in Physics Letters B. \href{https://doi.org/10.1016/j.physletb.2025.139302}{DOI: 10.1016/j.physletb.2025.139302} 

\vspace{2pt}

\textbf{Keywords:} Heavy-ion collision, Fusion mechanism, Dissipative dynamics, Langevin equation
\end{abstract}

\maketitle

\section{Introduction}

Over the past decades, remarkable progress has been made in the synthesis of superheavy nuclei; however, the underlying reaction mechanisms driving their production remain incompletely understood. The fusion mechanism in cold synthesis reactions, in particular the intermediate fusion phase, plays a pivotal role in the success of these processes. Dissipative dynamics frameworks, such as Langevin and Fokker-Planck equations, have emerged as powerful tools to address these challenges \cite{Feldmeier1987}. These approaches provide insights into experimental observables such as deep-inelastic cross-sections, energy dissipation, and mass and charge transfer \cite{Frobrich1996}. Furthermore, they have been successfully applied to quasifission, offering precise reconstructions of fission mass distributions \cite{Krappe2012}.

The Langevin approach provides an intuitive framework for nuclear dissipative dynamics, modeling phenomena such as heavy-ion reactions and nuclear fission as diffusion over a barrier \cite{Abe1986}. Early studies identified the role of the one-body dissipation and the fissility scaling in understanding the hindrance to compound nucleus formation \cite{Blocki1986}. These models laid the groundwork for stochastic approaches, which employed Langevin and Smoluchowski equations to capture the interplay between deterministic and stochastic forces in nuclear reactions \cite{Abe1996}.

Subsequent work applied these models to explore fusion-fission dynamics, emphasizing the critical role of asymmetry in the entrance channel and demonstrating that neutron-rich beams enhance fusion probabilities \cite{Aritomo1999, Abe1999}. A combined Langevin/statistical approach was developed to calculate fusion and evaporation residue cross sections and was validated on $^{48}$Ca-induced reactions \cite{Kosenko2002}. Later studies refined these methods, incorporating multidimensional stochastic descriptions and microscopic transport coefficients to align theoretical results with experimental data \cite{Litnevsky2014, Litnevsky2020}.

Recent studies using stochastic dynamics emphasized the interplay of shell effects, Coulomb barriers, and reaction Q-values in superheavy element formation. These insights guide the design of future experiments targeting new isotopes beyond $Z = 118$ \cite{Zagrebaev2012, Shen2011}.

This letter introduces a six-dimensional Langevin framework for heavy-ion fusion, which explicitly incorporates angular momentum, relaxes the constraint of a fixed mass asymmetry parameter, and includes all the off-diagonal elements of the relevant tensors, offering a significant improvement over the prior model \cite{Przystupa1994}.

\section{Shape parametrization}

Given the complexity of tracking all internal degrees of freedom during the fusion process, a common approach is to identify relevant collective degrees of freedom which effectively characterize the system. Within these frameworks, the slow collective degrees of freedom are viewed as being immersed in a bath of faster dynamics, representing the surrounding environment of individual particles. Friction and random forces arise naturally by defining an appropriate correlation function for the stochastic force governing interactions between the collective variables and the reservoir.

\begin{figure}[htb]
  \centering
  \includegraphics[width=0.7\columnwidth]{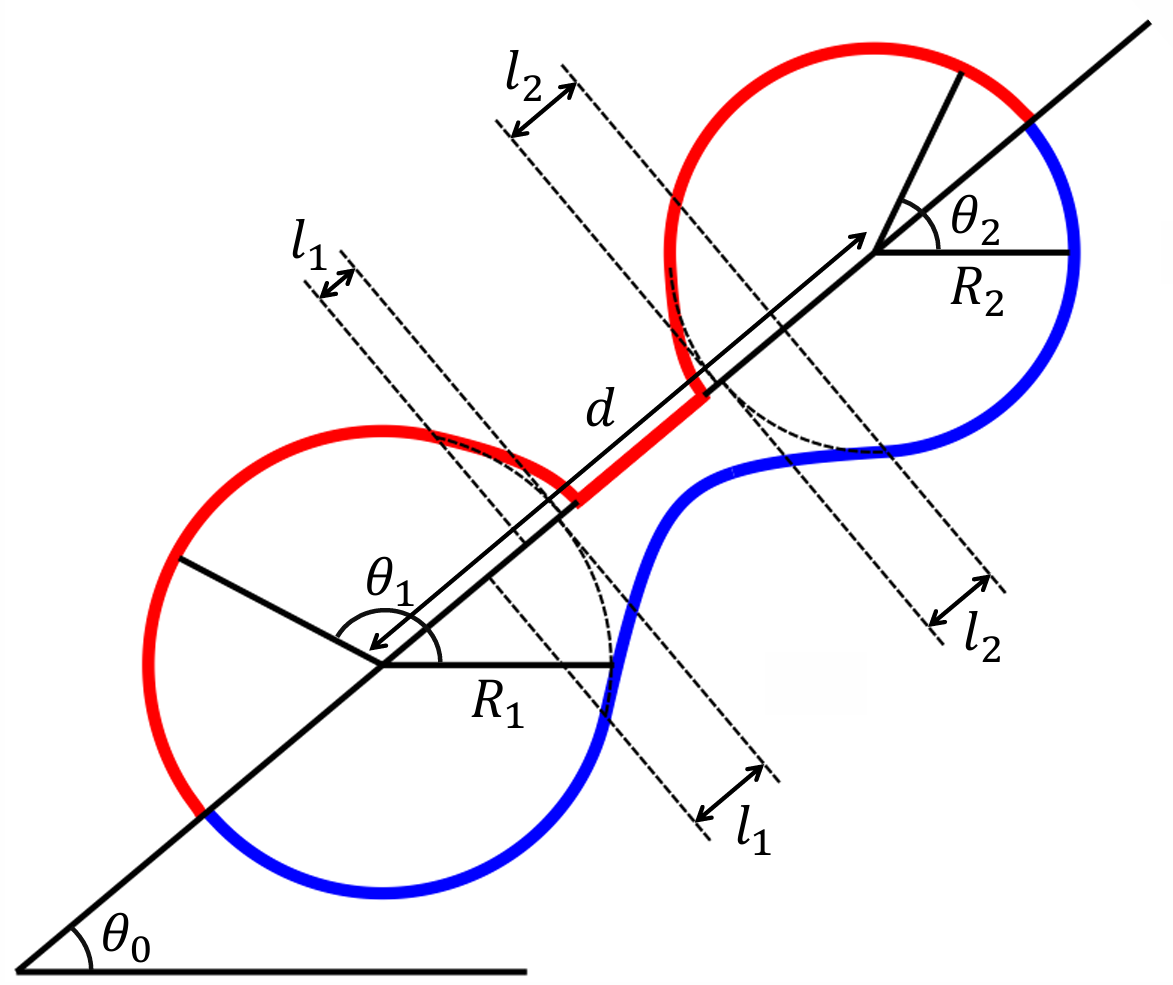}
  \caption{Examples of bipartite (top) and monopartite (bottom) shapes using the Błocki parametrization \cite{Blocki1982}. Please note that, as the total volume is constant, the radii\\ of the fragments in both shapes are different. For aesthetic reasons, distinct elongation and deformation variables were chosen for the configurations to match to the same distance $d$ between the fragments, but the asymmetry variable was kept the same.}
  \label{fig.parametrization}
\end{figure}

In Ref. \cite{Blocki1982}, Błocki and Świątecki introduced a parametri-zation for a system of two spherical ions, well suited for the description of cold fusion reactions. This parametrization captures both separated and monopartite shapes, maintaining a constant volume while progressively transitioning between shapes. Two sphere caps of different sizes form the axially symmetric shapes joined smoothly by a convoluted quadratic surface. The shapes are unequivocally defined by the radius $R_0$ of the associated spherical compound system, which acts as a scaling factor, and the three dimensionless coordinates:
\begin{equation}
\rho = \frac{d}{R_1+R_2}\, , \quad \lambda  = \frac{l_1+l_2}{R_1+R_2} \, , \quad \Delta = \frac{R_1-R_2}{R_1+R_2} \, ,
\end{equation}
namely the distance (or elongation), the neck (or deformation) and the asymmetry collective variables. $d$ denotes the distance between the centers of the spheres of radii $R_1$ and $R_2$ and $l_1$ and $l_2$, the distances from the innermost points of the spheres to their respective junctions with the quadratic surfaces (see Fig. \ref{fig.parametrization}). Unlike many other parametrizations, this approach offers the key benefit of precisely defining the scission moment, which occurs along the line $\lambda = 1 - 1/\rho$, making it particularly suited for the description of fusion and fission processes.

To investigate the impact of angular momentum dissipation and rotational effects, we additionally incorporate collective variables for rotational angles: the overall system angle is represented by $\theta_0$, while the angles for the first and second spheres are respectively denoted $\theta_1$ and $\theta_2$, as represented in Fig. \ref{fig.parametrization}.

\section{Langevin formalism} \label{Sec.Formalism}

Defining the vectors of collective variables $Q$ and their conjugated momenta $P$ as:
\begin{align}
Q &= (\rho,\lambda,\Delta,\theta_0, \theta_1, \theta_2)  , \\
P &= (p_\rho,p_\lambda,p_\Delta, \hbar \ell_0, \hbar \ell_1, \hbar \ell_2) \, ,
\end{align}
the Langevin system of equations reads:
\begin{align}
\dot{Q} &= \mathcal{M}^{-1}P, \label{eq.Langevin1}\\
\dot{P} &= -\nabla \mathcal{T} - \nabla \mathcal{V} -\Gamma \dot{Q} + GW. \label{eq.Langevin2}
\end{align}
For the shape degrees of freedom, the mass tensor $\mathcal{M}$ is determined in the incompressible and irrotational fluid approximation \cite{Davies1976}; in the rotational space, it is diagonal and its elements are the corresponding moments of inertia. The collective kinetic energy $\mathcal{T}=\frac{1}{2} \dot{Q}^{T}\!\!\mathcal{M}\dot{Q}$ is calculated from the mass tensor and includes all the $\dot{q}_i\dot{q}_j$ cross-terms in our calculations. The collective deformation potential $\mathcal{V}$ is obtained by subtracting the energy at the deformation $Q$ from the energy of the undeformed initial system. To ensure accurate excitation energies, $\mathcal{V}$ is adjusted by applying a shift equal to the difference $Q_{fus}^{exp} - Q_{fus}^{calc}$ between the calculated and experimental fusion $Q$-values. The $-\nabla(\mathcal{T}+~\mathcal{V})$ term thus represents the conservative forces acting on the macroscopic variables, while the remaining forces originate from the intrinsic degrees of freedom: the friction force $-\Gamma \dot{Q}$, which contributes to the irreversible production of heat, and the Langevin random force $GW$, which induces energy fluctuations. $\Gamma$ and $G$ stand for the friction and fluctuation tensors, and $W$ a vector of random distributions discussed in detail further in the text. The shape friction is calculated using the wall-plus-window friction model \cite{Blocki1978,Blocki1977}, in which a smooth transition between the mononuclear and dinuclear dissipation regimes was implemented following Ref. \cite{Feldmeier1987}. We also assume the proximity formalism \cite{Blocki1978, Blocki1977} via the definition of an effective window opening, larger than the geometrical one. This proximity formalism  allows the fragments to undergo friction even without direct contact and thus accounts for quantum effects, such as wavefunction tail effects or quantum tunneling. In our calculations, we use the universal proximity distance $s_{prox} = 3.2$~fm, which indicates the distance at which the fragment surfaces are close enough for the fragments to interact \cite{Blocki1977}. The angular friction consists of the sliding friction as given in Refs. \cite{Tsang1974,Fai1983}; the rolling part which is an order of magnitude smaller is neglected \cite{Tsang1974}. The fluctuation tensor $G$ appearing in the Langevin force satisfies $G^2 = D$, where $D$ is the diffusion tensor governed by the Einstein relation $D = T^*\,\Gamma $. The quantum-corrected temperature $T^*$ is related to the classical temperature $T$ through $T^*=E_0/\tanh(E_0/T)$ \cite{Hofmann1977, Hasse1979}, with $E_0$ being the zero-point collective energy of the oscillators forming the heat bath. For surface oscillations, $E_0$ was estimated as 0.45 MeV (quadrupole/elongation) and 2.23 MeV (hexadecapole/neck)~\cite{HillWheeler1953}. Recent studies \cite{Pomorski2023, Ivanyuk2021} fit $E_0$ to experimental data, yielding a standard value of $E_0=1-1.5$ MeV. Here, we adopt a higher value $E_0 = 2$ MeV to account for the significant neck increase at our initial condition, resembling the hexadecapole mode from Ref. \cite{HillWheeler1953}. The classical temperature $T$ is calculated from the internal excitation energy $E^*$ along each trajectory, using the formula $T=\sqrt{E^*/a}$, in which the level density parameter $a$ is given by $a=A/n $ MeV$^{-1}$. Standard values $n = 8-12$ yield similar results beyond $T = 2$ MeV \cite{Pomorska2006}, which is our choice for $E_0$, and we adopt $n=8$ in this study.

\begin{figure*}[ht]
  \centering
  \includegraphics[width=0.95\textwidth]{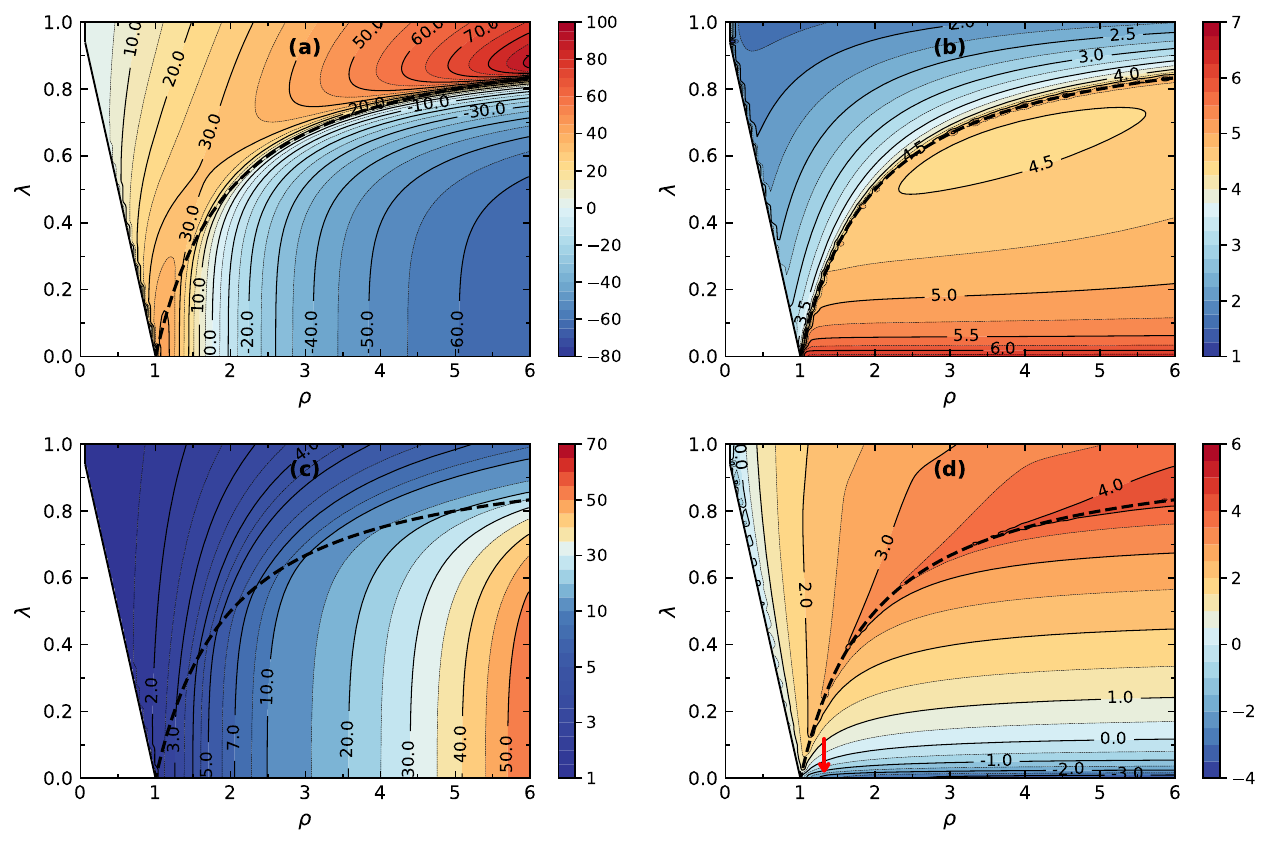}
  \vspace{-10pt}
  \caption{The panels (a)-(d) display selected physical quantities at $\Delta =\Delta_{init}$ for the $^{64}$Ni + $^{92}$Zr system: $\mathcal{V} + Q_{fus}$ [MeV] (a), $\log(\mathcal{M}_{\Delta \Delta} / (1\hbar^2\!$/MeV$))$ (b), $I_{tot} / I_{tot}^{sph}$ in a pseudo-logarithmic scale (c) and $\log(\Gamma_{\lambda \lambda} / (1\hbar))$ (d). The white regions on the left represent forbidden configuration space~\cite{Blocki1982}, while the dashed line indicates the scission line, separating the monopartite regime (above) from the bipartite regime (below). Proximity friction begins at ($\rho \simeq 1.32$, $\lambda = 0$) preceding contact (red arrow on the panel (d)).}
  \label{fig.ingredients}
\end{figure*}

The Langevin equations, as presented in Eqs.~(\ref{eq.Langevin1}, \ref{eq.Langevin2}), treat the shape and the angular variables on the same footing. However, ensuring the angular momentum conservation $\ell_0 + \ell_1 + \ell_2 = L_{tot}$, requires addressing conflicts with the randomness introduced by the Langevin force. While independent angular variables can be approximately extracted in exceptional cases \cite{Tsang1974,Fai1983}, or expressed via the estimation of a mean tangent friction in bipartite systems \cite{Amano2022}, our formalism accommodates both monopartite and bipartite regimes. In this work, each component (fragments 1, 2, and the neck) is allowed to evolve freely during a given integration step of the Langevin equations, after which a corrective factor $L_{tot}/\sum \ell_i$ is applied to each angular momentum to enforce angular momentum conservation.

In this introductory paper, with the aim of understanding the general features of the Langevin approach, simple potentials and random forces were selected. The nuclear potential is modeled using a Yukawa-plus-exponential folding potential with standard parameters taken from Ref. \cite{Krappe1979}. The Coulomb potential energy is calculated under the assumption of a uniform charge distribution \cite{Feldmeier1987}. Similarly, the random force is modeled as a Gaussian (white) noise characterized by $\langle W(t)\rangle = 0$, $\langle W(t)W(t')\rangle = 2\delta(t-t')$. The Langevin equations are discretized and numerically integrated with a time step $\tau = 10^{-25}$ s, and the collective variables $Q$ are integrated using the Heun method $Q_{n+1} = Q_{n} + \left[\mathcal{M}^{-1}\right]_n \frac{P_n+P_{n+1}}{2}\tau$, enhancing the precision quadratically. It is also worth noting that the discretization of the time-dependent random forces leads to $\int_{t_n}^{t_n+\tau} W(t)dt \simeq \sqrt{\tau}W_n$, where $W_n$ are vectors of random numbers drawn, at each iteration $n$, from a normal distribution of variance 2.

Figure \ref{fig.ingredients} presents a selection of physical quantities pertinent to the Langevin formalism for the $^{64}$Ni + $^{92}$Zr system, with the asymmetry variable set to its value for the undeformed separated system $\Delta =\Delta_{init}$. In all panels, the scission line, which separates the monopartite and the bipartite regimes is indicated by a dashed line. In Fig. \ref{fig.ingredients}(a), the quantity $\mathcal{V} + Q_{fus}$ is displayed. The spherical compound state (top left corner of the graph) is associated to the value zero, while at large distances, in the absence of deformation, the potential approaches $Q_{fus} =-91.7$ MeV. In the intermediate stage, the potential barrier emerges as the fragments come into contact. Figure \ref{fig.ingredients}(b) provides an example of mass tensor element, specifically the logarithm of the diagonal element in the asymmetry variable $\mathcal{M}_{\Delta \Delta}$ in $\hbar^2\!/$MeV units. The plot clearly delineates the separation between the monopartite and the separate regimes. In the separate regime, $\mathcal{M}_{\Delta \Delta}$ reaches extremely high values, illustrating the absence of matter exchange between the fragments. The total moment of inertia about the center of mass is shown in Fig. \ref{fig.ingredients}(c) in the form of $I_{tot} / I_{tot}^{sph}$ in a pseudo-logarithmic scale. This graph illustrates the transition from high values when the fragments are separated to the spherical compound nucleus value $I_{tot}^{sph} = 60.2$ $\hbar^2$/MeV as the distance decreases. Finally, Figure \ref{fig.ingredients}(d) depicts the logarithm of the friction matrix element $\Gamma_{\lambda \lambda}$ in $\hbar$ units. Although the smoothing procedure applied by the plotting software may diminish the apparent strength of the proximity effect, one can notice the increase of friction prior to contact, at the elongation $\rho_{prox} \simeq 1.32$ (red arrow).

To speed up the calculations, the potential, mass, and friction tensor elements are precomputed on a $100 \times 100 \times 100$ grid spanning $\rho \in [0, 3], \lambda \in [0, 1], \Delta \in [0, 1]$. Their values and derivatives are extrapolated along each trajectory using cubic B-splines, with five control points per dimension. Careful attention is given to selecting appropriate extrapolation regions, which should remain within the geometrical configuration space and the relevant regime (monopartite or bipartite, with or without proximity) to ensure proper extrapolation.

\section{Fusion mechanism}

In the case of heavy colliding nuclei, surpassing the Coulomb barrier alone does not guarantee fusion into a compound nucleus, as the repulsive Coulomb forces counteract the nuclear forces upon contact. The Langevin equations reveal that the fusion mechanism unfolds in three distinct phases summarized in Fig. \ref{fig.fusion}.

\subsection{Initial deceleration phase}

\begin{figure}[htb]
  \centering
  \includegraphics[width=0.95\columnwidth]{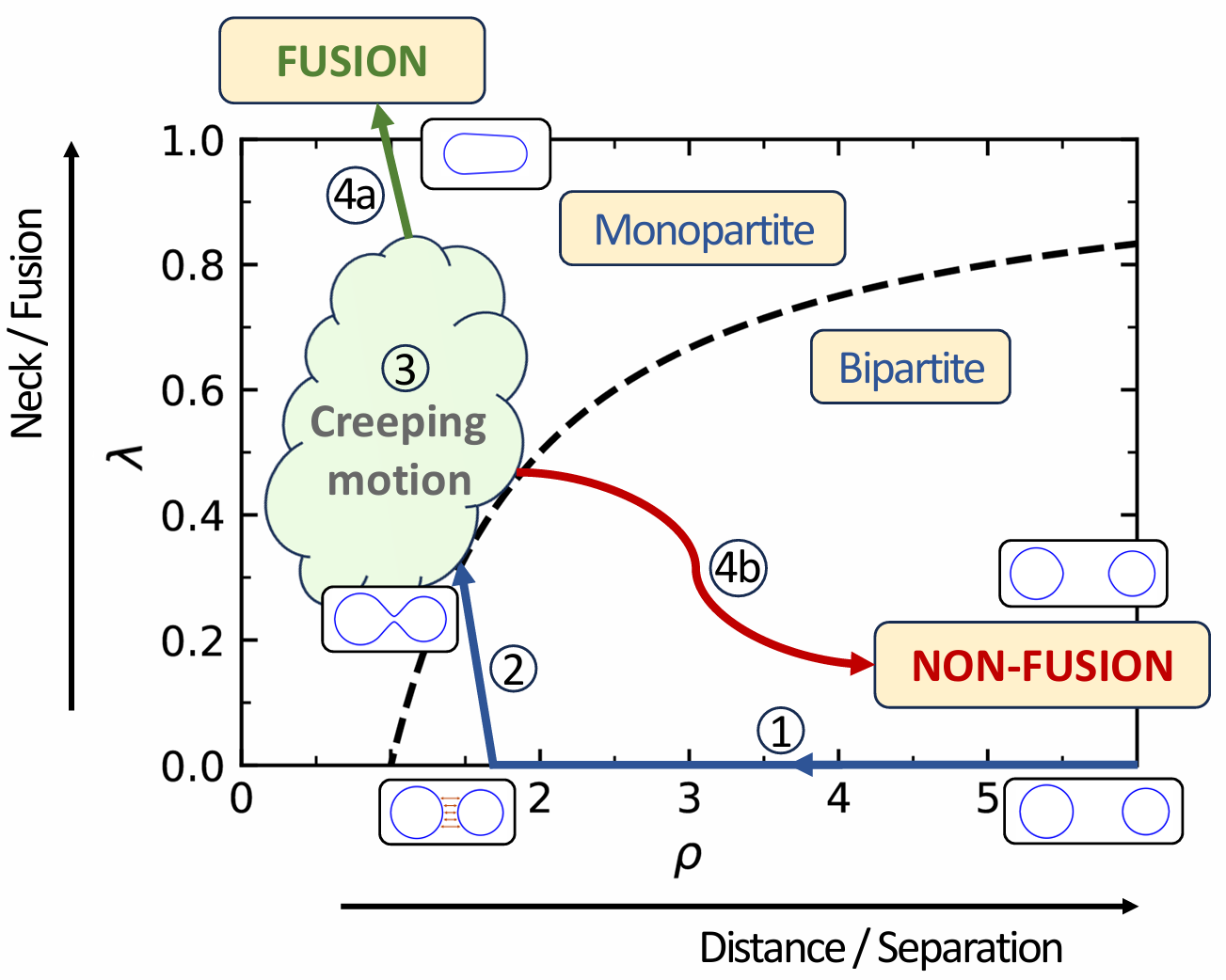}
  \caption{Schematic representation of the fusion stages. The dashed line marks the scission line, which divides the monopartite and bipartite regimes.}
  \label{fig.fusion}
\end{figure}

During the initial stage (Fig. \ref{fig.fusion}, label 1), the system undergoes rapid deceleration, losing a significant portion of its kinetic energy. This phase is characterized by near-zero deformation, as reflected in the Langevin equation for the neck velocity\footnote{Without loss of generality, the asymmetry terms are omitted for the purposes of the discussion.}:
\begin{equation}
\dot{\lambda}=\mathcal{M}^{-1}_{12}p_{\rho}+\mathcal{M}^{-1}_{22}p_{\lambda} \, . \label{eq.neckVelocity}
\end{equation}
The inverse mass tensor cross element $\mathcal{M}^{-1}_{12}$, which is positive, indicates that a decrease in elongation naturally reduces deformation geometrically. At the $\lambda = 0$ boundary, as the fragments approach, the negative elongation momentum $p_\rho$ causes the first term $\mathcal{M}^{-1}_{12}p_{\rho}$ to tend towards $-\infty$, effectively suppressing any deformation. The system remains close to the border, where the second term $\mathcal{M}^{-1}_{22}p_{\lambda}$ tends to + $\infty$ to act as a counterbalance to the first term and prevent the crossing of the boundary: the neck velocity $\dot{\lambda}$ remains zero and the fragments undeformed. This behavior might seem surprising, but is sensible from a physical point of view. Any additional deformation of the fragments requires the two fragments to interact (or have interacted) via the nuclear force. Such an interaction, apart from long-range Coulomb effects, has yet to occur during the initial phase. The Langevin equations are integrated directly in this first stage, as only conservative forces are involved.

\subsection{Contact phase}

The deceleration phase concludes when the fragments begin to interact, which, in our model, occurs before contact at a distance $d_{prox} = R_{1,0} + R_{2,0} + s_{prox}$, where $s_{prox} = 3.2$~fm is the previously defined proximity separation, and $R_{i,0}$ the initial radii of the fragments. The emergence of friction slightly reduces the elongation momentum, which disrupts the balance of infinities. The $\mathcal{M}^{-1}_{22}p_{\lambda} =+\infty$ term then dominates the neck velocity in Eq. (\ref{eq.neckVelocity}), leading to an abrupt increase of the neck velocity to $+\infty$. The system is propelled to the scission line, where the fragments make contact (Fig. \ref{fig.fusion}, label 2).

This phenomenon, often called the "sudden approximation" in mean-field formalisms \cite{Pomorski1980}, stems from the inherent instability of non-saturated nucleonic densities. Such anomalous densities occur when the tails of the nuclear wavefunctions overlap and the summed densities reach the saturated value. At this point, contact, in macroscopic terms, occurs almost instantaneously. In our model, estimating the push in the neck direction is challenging, as it occurs at the $\lambda = 0$ boundary where physical quantities diverge. We thus assume an instantaneous transition to the scission line. The entrance point is considered the same, regardless of the initial radial kinetic energy. This is a reasonable approximation since larger elongation momenta also imply stronger pushes in the neck direction. Geometrical considerations of deformation and approximations to the mass tensor near $\lambda = 0$ allow us to estimate this contact point at approximately $(\rho_0, \lambda_0, \Delta_0)$, where $\rho_0 = d_0/(R_{1,0} + R_{2,0})$, $d_0 = R_{1,0} + R_{2,0} + 2.0$ fm, $\lambda_0 = 1 + 1/\rho_0$ and $\Delta_0=\Delta_{init}$. Due to the high friction matrix element values $\Gamma_{\lambda\lambda}$ at the scission line caused by the regime change (see Fig. \ref{fig.ingredients}(d)), the system enters the monopartite regime with negligible neck and asymmetry velocities. Assuming this transition is rapid and has minimal impact on the elongation momentum, the Langevin equations Eq. (\ref{eq.Langevin2}) for the neck and asymmetry variables are inverted to determine their initial momenta for the third fusion stage.

\subsection{Creeping motion phase}

Following contact, the system enters a prolonged creeping motion phase, where the nuclear and Coulomb forces oppose each other (Fig. \ref{fig.fusion}, label 3). The Langevin equations are solved numerically with the initial conditions estimated in the previous sections. The process ultimately leads to one of two outcomes: fusion or separation of the nuclei, marked as 4a and 4b in Figure \ref{fig.fusion}, respectively.

In our calculations, we verified that the conditions $\lambda = 1$ (half-sphere mixture), $\rho(1-\lambda) = \Delta^2$ (window angle fully open \citep{Blocki1982}), or $\rho = 0.5$ (small distance) consistently lead to the fusion of the system. As a result, we terminate the calculations when any of these conditions are met. Conversely, when $\lambda = 10^{-2}$ or $\rho = 3$, we consider that the system is heading towards separation and we also end the trajectory.

\section{Spin distribution and fusion cross section}

Due to the presence of the fluctuation forces, the resolution of the Langevin equations generates a distribution of trajectories. The initial angular momentum is also assigned randomly with $\ell_{init} = \ell_{max}\sqrt{x}$, where $x$ is a random variable distributed uniformly in $[0;1]$. This method ensures a linear progression of the initial momenta from $0$ to $\ell_{max}$. The spin distribution for a specific bin $i$  is then computed as a Monte-Carlo integral following the equation:
\begin{equation}
\sigma_{\ell_i} \equiv \left(\frac{\d \sigma_{\! fus}}{\d \ell} \right)_{\!\ell_i} \!\! = \frac{2\pi}{k^2}\ell_i\frac{N_i^{fus}}{N_i^{tot}},
\end{equation}
where $\ell_i$ is the angular momentum of the bin $i$, $N_i^{fus}$ and $N_i^{tot}$ are respectively the corresponding number of fusing and total number of trajectories, and $k = \sqrt{2\mu E_{cm} / \hbar ^2}$ with $\mu$ the reduced mass of the system, and $E_{cm}$ the energy in the center of mass coordinates. The maximum angular momentum $\ell_{max}$ is chosen empirically at a value where the partial cross-sections are assumed to have dropped to zero. The total fusion cross section $\sigma_{\! fus}$ is then obtained by integrating the spin distribution across all bins.

\section{Results and discussion}

The model was applied to the systems $^{64}$Ni + $^{92}$Zr and $^{64}$Ni + $^{96}$Zr, for which the projectiles are mostly spherical in their ground states. $^{96}$Zr, being neutron-rich, may exhibit a slight deviation from sphericity. Experimental measurements of total fusion cross-sections were reported in Refs. \cite{Kuhn1989, Stefanini1990} at the energies $E_{cm}=138.8$ MeV and $E_{cm}=139.5$ MeV, respectively. These energies correspond to excitation energies of approximately 50 MeV, where shell effects are negligible. Spin distributions were extracted in Ref. \cite{Kuhn1989}, though these model-dependent results are subject to uncertainties, in particular at low angular momenta.

A previous study of these reactions using Langevin formalism (Ref. \cite{Przystupa1994}) employed the same collective variables as in this work. Computational constraints at the time required the asymmetry variable to be fixed, which was believed to explain the sharp tails observed in the spin distributions. However, the present study highlights the importance of properly treating configuration boundaries, regime changes, and separating the fusion process into three distinct stages to achieve physical asymptotic spin distributions.

\begin{figure}[htb]
  \center
  \includegraphics[width=0.90\columnwidth]{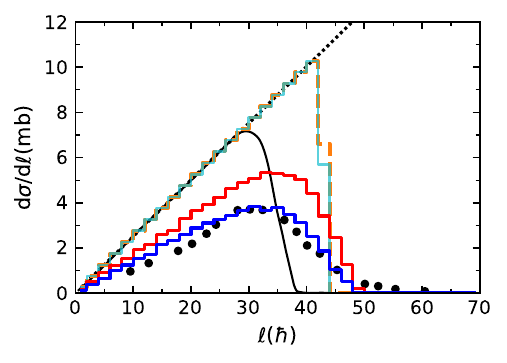}
  \includegraphics[width=0.90\columnwidth]{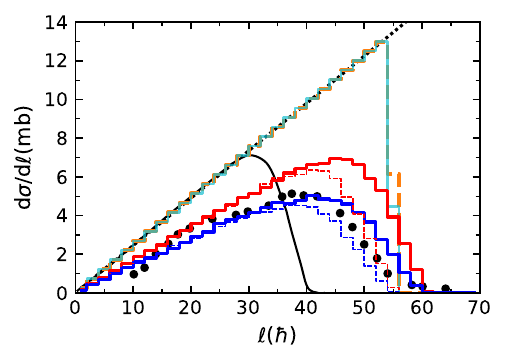}
  \vspace{-10pt}
  \caption{Fusion differential cross sections for $^{64}$Ni+$^{92}$Zr at $E_{cm}$ = 138.8 MeV (top) and $^{64}$Ni+$^{96}$Zr at $E_{cm}$ = 139.5 MeV (bottom). The top dashed line represents the unitarity limit. The orange dashed curve (no Langevin force) and the red curve (with Langevin forces) correspond to calculations with a frozen asymmetry parameter. The cyan and solid blue curves are the corresponding results in the full configuration space. For the second system, dashed thin curves represent results without the potential energy correction. Experimental data, shown as black dots, are taken from Refs. \cite{Kuhn1989, Stefanini1990}. The solid black curve represents the raw calculations from Ref. \cite{Przystupa1994}.}
  \label{fig.spinDistributions}
\end{figure}

Using the Yukawa-plus-exponential folding + Coulomb potential previously described, the calculated fusion Q-values are $Q_{fus}=-91.7$ MeV (experimental: $-91.4$ MeV) for $^{64}$Ni + $^{92}$Zr and $Q_{fus}=-88.9$ MeV (experimental: $-86.5$ MeV) for $^{64}$Ni + $^{96}$Zr. For the latter system, the 2.4 MeV energy shift discussed in Sec. \ref{Sec.Formalism} ensures accurate energetics.

Figure \ref{fig.spinDistributions} presents the resulting spin distributions. The top dashed line indicates the unitarity/maximal limit of differential cross-sections. The black dashed curve shows the raw spin distributions from Ref. \cite{Przystupa1994}, exhibiting the sharp asymptotic behavior. The orange and cyan curves correspond to calculations without the Langevin forces, with fixed and free asymmetry, respectively, displaying the typical abrupt cutoffs at high angular momentum. In the present cases, the cutoffs happen within the same $\ell$-bin, causing the two curves to overlap everywhere except within that particular bin. A more gradual asymptotic behavior is only seen with the incorporation of the Langevin forces which introduce energy fluctuations. The red curves show the results with the asymmetry fixed (for comparison to the calculations of Ref. \cite{Przystupa1994}), while the blue curves are our final results in the full six-dimensional space. Distributions without the 2.4 MeV energy shift are shown as dashed thin lines for the $^{64}$Ni + $^{96}$Zr system.

In the $^{92}$Zr case, the spin distributions closely match experimental data, as well as the total fusion cross-section $\sigma_{fus} = 107$ mb (experimental: 100 mb). In the case of $^{96}$Zr, there is a slight discrepancy with experimental data, potentially due to $^{96}$Zr deformation, as evidenced by the double-peak structure in the experimental distributions. However, the energy shift aligns the calculated total cross-section $\sigma_{fus} = 175$ mb with the experimental value (166 mb) compared to $\sigma_{fus} = 145$ mb without the shift.

\section{Conclusions and perspectives}

In this study, we have developed a six-dimensional Langevin-based formalism to investigate the fusion mechanism of heavy ions, incorporating elongation, neck and asymmetry variables with unrestricted motion. The dynamics naturally lead to overdamping and rapid neck stabilization. Using a simple Yukawa-plus-exponential folding potential and Gaussian random forces, the resulting spin distributions and fusion cross-sections exhibit excellent agreement with experimental data for the systems $^{64}$Ni + $^{92, 96}$Zr at excitation energies of approximately 50 MeV, offering a robust validation of the model. The study paves the way for exploring diverse phenomena such as reaction times, mass-angle distributions (MADs), quasifission, and angular momentum dissipation by providing a direct access to key variables and their interplay.

Future efforts will focus on various enhancements. To achieve a fully microscopic-macroscopic representation, we plan to incorporate shell effects, allowing studies of fusion at lower excitation energies. Additionally, we aim to explore various forms of stochastic colored noise to investigate memory effects and their implications in the fusion process. These advancements will further broaden the applicability of the model, enabling deeper insights into fusion dynamics, the synthesis of superheavy elements and addressing the challenging hindrance mechanism in heavy-ion fusion.

\section*{ACKNOWLEDGMENTS}

We gratefully acknowledge David Boilley for engaging in numerous discussions. We also extend our gratitude to the CIŚ computing center at the National Centre for Nuclear Research for providing extensive computation resources essential to this study. M.K. was partially co-financed by the COPIGAL Project.

\bibliographystyle{apsrev4-2}  
\bibliography{Langevin_bibfile}

\end{document}